\documentstyle[twocolumn,aps,epsf]{revtex}

\begin{document}
\draft                              

\title{
Magnetic-field-induced singularities in spin dependent 
tunneling through InAs quantum dots}

\author{I.~Hapke-Wurst,$^1$ U.~Zeitler,$^1$  
H.~Frahm,$^2$ A.~G.~M.~Jansen,$^3$ R.~J.~Haug,$^1$ and K.~Pierz$^4$}

\address{
$^1$Institut f\"ur Festk\"orperphysik, Universit\"at Hannover, 
Appelstra{\ss}e 2, D-30167 Hannover, Germany\\
$^2$Institut f\"ur Theoretische Physik, Universit\"at Hannover, 
Appelstra{\ss}e 2, D-30167 Hannover, Germany\\
$^3$Grenoble High Magnetic Field Laboratory, MPIF-CNRS, B.P.~166,  
F-38042 Grenoble Cedex 09, France\\
$^4$Physikalisch-Technische Bundesanstalt Braunschweig,
Bundesallee 100, D-38116 Braunschweig, Germany} 

\date{\today}
\maketitle

\begin{abstract}
Current steps attributed to resonant tunneling through individual 
InAs quantum dots embedded in a GaAs-AlAs-GaAs tunneling device are 
investigated experimentally in magnetic fields up to 28~T.
The steps evolve into strongly enhanced current peaks in high fields.
This can be understood as a field-induced Fermi-edge singularity due
to the Coulomb interaction between the tunneling electron on the
quantum dot and the partly spin polarized Fermi sea in the Landau quantized 
three-dimensional emitter.  
\end{abstract}

\narrowtext

The interaction of the Fermi sea of a metallic system with a local 
potential can lead to strong singularities close to the Fermi edge.
Such effects have been predicted more than thirty years ago 
for the X-ray absorption and emission of metals\cite{theoXray} and observed 
subsequently\cite{expXray}.
Similar singularities as a consequence of many body effects
are also known from the luminescence of quantum wells\cite{Lum}.
Matveev and Larkin were the first to predict interaction-induced 
singularities in the tunneling current via a localized 
state\cite{Matveev:1992} which were measured experimentally in 
several resonant tunneling
experiments\cite{Geim:1994,Cobden:1995,Benedict:1998}
from {\sl two-dimensional} electrodes through a zero-dimensional system.

Here we report on singularities observed in the resonant tunneling from 
highly doped {\sl three-dimensional} (3D) GaAs electrodes through an InAs 
quantum dot (QD) embedded in an AlAs barrier. 
These Fermi-edge singularities (FES) show a considerable magnetic field 
dependence and a strong enhancement in high 
magnetic fields where the 3D electrons occupy the 
lowest Landau level in the emitter. 
We observe an asymmetry in the enhancement for electrons of different 
spins with an extremely strong FES for electrons carrying the majority 
spin of the emitter.
The experimental observations are explained by a theoretical model 
taking into account the electrostatic potential experienced by the 
conduction electrons in the emitter due to the charged QD.
We will show that the partial spin polarisation of the emitter causes 
extreme values of the edge exponent $\gamma$ not observed until present 
and going beyond the standard theory valid for 
$\gamma \ll 1$~\cite{Matveev:1992}.

The active part of our samples are self-organized InAs QDs with 3-4~nm height 
and 10-15~nm diameter embedded in the middle of a 10~nm-thick AlAs barrier and 
\begin{figure}
\vspace*{1em}
\centerline{\epsfxsize=7cm 
\epsfbox{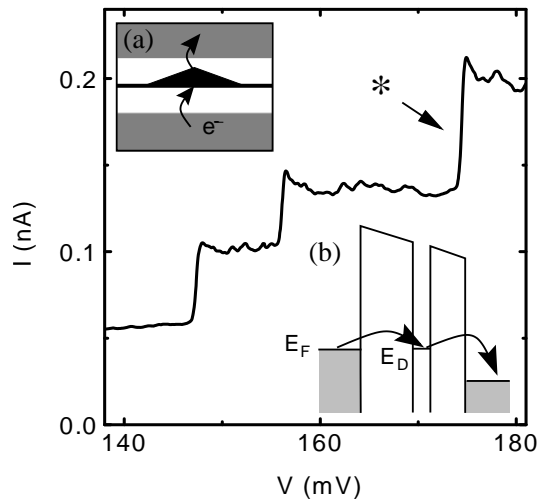}}
\vspace*{1em}
\caption{Typical steps in the $I$-$V$-characteristics of a GaAs-AlAs-GaAs
tunneling diode containing InAs QDs at T~=~500~mK. The (*) marks
the current step due to single electon tunneling through one individual
InAs QD which is analyzed in detail.
Insets: (a) Principle sample structure of an InAs QD (black) embedded 
in an AlAs-barrier (white) between two GaAs-electrodes (grey). 
The arrows mark the tunneling direction of the electrons. 
(b) Schematic profile of the band structure at positive bias.}
\label{steps}
\vspace*{1em} 
\end{figure}
sandwiched between two 3D electrodes.
They consist of a 15~nm undoped GaAs spacer layer and a GaAs-buffer with 
graded doping.
A typical InAs dot is sketched in inset (a) of Fig.~\ref{steps}, the vertical 
band structure across a dot is schematically shown in inset (b).

Current voltage ($I$-$V$) characteristics were measured in large area vertical 
diodes ($40\times 40~\mu$m$^2$) patterned on the wafer.
In Fig.~\ref{steps} we show a part of a typical $I$-$V$-curve 
with several discrete steps.
We have demonstrated previously  that such steps can be assigned to single 
electron tunneling from 3D electrodes through individual InAs QDs 
\cite{Hapke:1999} consistent with other resonant tunneling experiments through 
self-organized InAs QDs~\cite{tunnel}. 

For the positive bias voltages shown in Fig.~\ref{steps} the electrons tunnel 
from the bottom electrode into the base of an InAs QD and leave the dot via 
the top.
The tunneling current is mainly determined by the tunneling rate through the 
effectively thicker barrier below the dot (single electron tunneling regime).
A step in the current occurs at bias voltages where the energy level of a dot, 
$E_D$, coincides with the Fermi level of the emitter, $E_F$. 

In the following we will concentrate on the step labeled (*) in 
Fig.~\ref{steps}. Other steps in the same structure as well as steps 
observed in the $I$-$V$-characteristics of other structures show a very 
similar behavior.

After the step edge a slight overshoot in the tunneling current occurs
consistent with other tunneling experiments through a localized 
impurity~\cite{Geim:1994} or through InAs dots~\cite{Benedict:1998}.
This effect is caused by the Coulomb interaction between a localized 
electron on the dot and the electrons at the Fermi edge of the emitter.
The decrease of the current $I(V)$ towards higher voltages
$V >V_0$ follows a power law 
$I \propto (V-V_0)^{-\gamma}$~\cite{Geim:1994} 
($V_0$ is the voltage at the step edge) with an edge exponent 
$\gamma = 0.02 \pm 0.01$.

\begin{figure}
\vspace*{1em}
\centerline{\epsfxsize=8cm 
\epsfbox{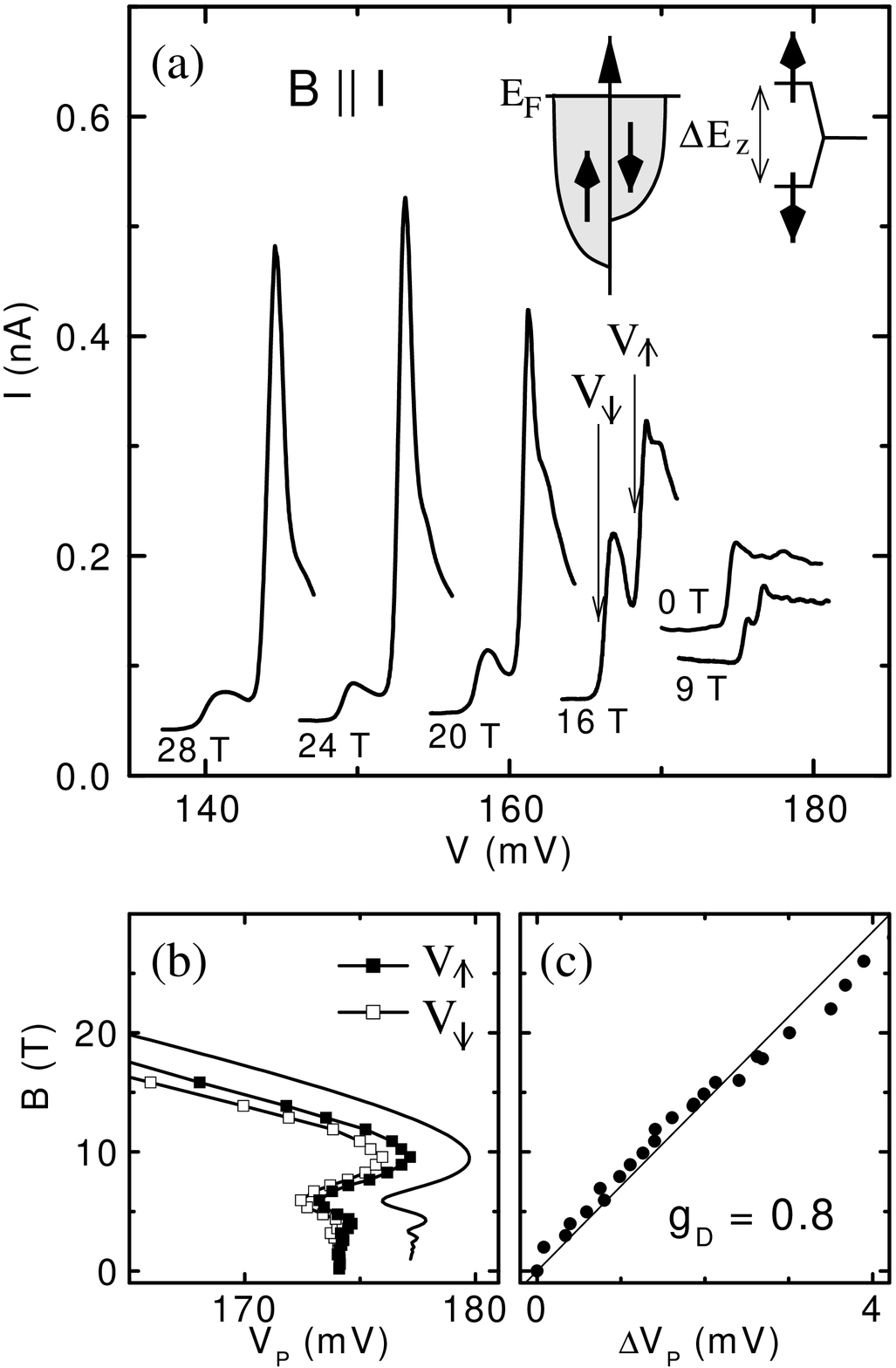}} 
\vspace*{2em} 
\caption{
(a) $I$-$V$-characteristics of step (*) at T~=~500~mK 
in various magnetic fields 
up to 28 T. The inset sketches the partial spin polarization 
in the emitter and the spin splitting of the dot level in a magnetic field.\\ 
(b) Onset voltages for the two spin-split 
current steps as a function of $B$
compared to the expected behavior for $E_F = 13.6$~meV and
$\Gamma= 1.3$~meV (solid line, shifted to the right for clarity).\\  
(c) Measured Zeeman splitting between the two current steps.}

\label{Babh}
\vspace*{1em}
\end{figure}

The evolution of step (*) in a magnetic field applied parallel to the 
current direction is shown in Fig.~\ref{Babh}a.
The step develops into two separate peaks with onset voltages marked as 
$V_\downarrow$ and $V_\uparrow$.
The Landau quantization of the emitter leads to an oscillation
of $V_\downarrow$ and $V_\uparrow$ and a shift to smaller voltages as a 
function of magnetic field, see Fig.~\ref{Babh}b.  
This reflects the magneto-quantum-oscillation of the Fermi energy in 
the emitter~\cite{Bumbel,Main:2000}. 
From the period and the amplitude of the oscillation we can extract
a Fermi energy (at $B = 0$) $E_0 = 13.6~$meV and a Landau level 
broadening $\Gamma = 1.3~$meV in the 3D emitter. The measured 
$E_0 = 13.6~$meV agrees well with the expected electron concentration 
at the barrier derived from the doping profile in the electrodes.

For $B > 6$~T only the lowest Landau level remains occupied.
With a level broadening $\Gamma = 1.3~$meV the Fermi level $E_F$
for 15 T $<$ B $<$ 30 T is within less than $2~$meV pinned to the bottom 
of the lowest Landau band, $E_L = \hbar \omega_c/2$.  
As a consequence the onset voltage shifts to lower values as 
$\alpha e\Delta V \approx -\hbar \omega_c/2$ with $\alpha = 0.34$.
The diamagnetic shift of the energy level in the dot 
can be neglected compared to this shift of the Fermi energy in the emitter. 
For the dot investigated in \cite{Hapke:1999} with $r_0 = 3.7$~nm the 
diamagnetic shift at 30~T is $\Delta E_D = 3.5$~meV negligible compared to 
$E_L = 26~$meV.

The two distinct steps with onset voltages $V_\downarrow$ and $V_\uparrow$
originate from the spin-splitting of the energy level $E_D$ in the dot.
Their distance $\Delta V_p$ is given by the Zeeman splitting
$\Delta E_z = g_D \mu_B B = \alpha e \Delta V_p$ 
with an energy-to-voltage conversion factor 
$\alpha = 0.34$~\cite{explain-alpha}.
As shown in Fig.~\ref{Babh}c $\Delta V_p$  is indeed linear in B, 
with a Land\'e factor $g_D = 0.8$ in agreement with other 
experiments on InAs dots~\cite{Thornton:1998}.

 For low magnetic fields ($B \le 9~$T in our case, see graph  for $B = 9~T$
in Fig.~\ref{Babh}a) the size of the steps is very similar for both spins 
and about half of the size at zero field.
Also the slight overshoot in the current as the signature of a Fermi 
edge singularity is similar for both spin orientations and comparable to 
the zero field case with an edge exponent $\gamma < 0.05$ 
for all magnetic fields $B < 10~$T.


The form of the current steps changes drastically in high magnetic fields 
where only the lowest Landau level of the emitter remains occupied.
In particular, the second current step at higher voltage evolves 
into a strongly enhanced peak with a peak current of one order of magnitude
higher compared to the zero-field case.

\begin{figure}
\vspace*{1em}
\centerline{\epsfxsize=8cm 
\epsfbox{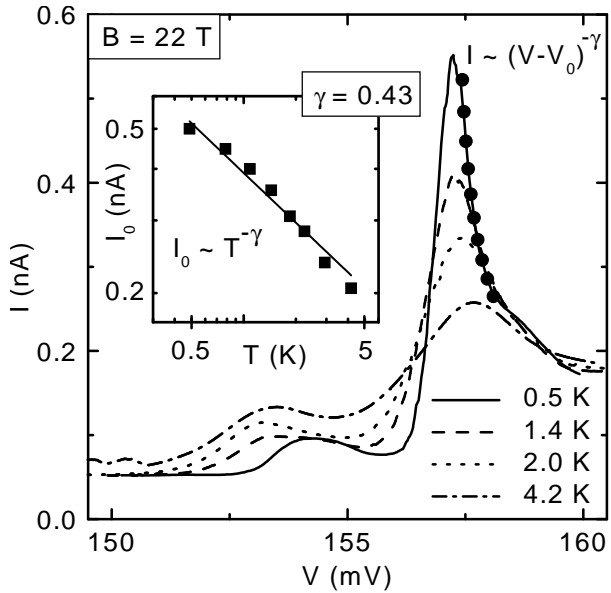}} 
\vspace*{2em} 
\caption{$I$-$V$-characteristics at B = 22 T for different temperatures.
The dots represent the shape of a FES,
$I \propto (V-V_0)^{-\gamma}$, with an edge exponent $\gamma = 0.43$. 
The inset shows the peak height as a function of temperature with
the solid line fitting the peak current $I_0$ by a power law 
$I_0 \propto T^{-\gamma}$.}  
\label{Tabh}
\vspace*{1em}
\end{figure}

Following~\cite{Thornton:1998} we assume that $g_D$ is positive whereas 
the Land\'e factor in bulk GaAs is negative.
This assumption is verified by the fact that the energetically lower lying 
state (first peak in Fig.~\ref{Tabh}) is thermally occupied at higher 
temperatures and can therefore be identified with the minority spin in the 
emitter. 
The strongly enhanced current peak at higher energies is due 
to tunneling through the spin state corresponding to the majority spin 
(spin up) in the emitter. The resulting spin configuration is scetched in 
the inset of Fig.~\ref{Babh}a and will also be confirmed below by our 
theoretical results. 

The shape of this current peak can be described by a steep ascent 
and a more moderate decrease of the current towards higher voltages. 
Down to temperatures $T<100$~mK the steepness of the ascent is only 
limited by thermal broadening. 
The decrease of the current for $V >V_0$ is again described with 
the characteristic behavior for a Fermi-edge singularity, 
$I \propto (V-V_0)^{-\gamma}$, where $V_0$ here is the voltage at 
the maximum peak current.

However, along with the drastic increase of the peak current
the edge exponent $\gamma$ increases dramatically 
reaching a value $\gamma > 0.5$ for the highest fields.


A different way to visualize the signature of a FES
is a temperature dependent experiment. As an example we have plotted the  
$I$-$V$-curve at $B=22$~T for different temperatures in Fig.~\ref{Tabh}. 
As shown in the inset the peak maximum $I_0$ for the spin-up electrons
decreases according to a power law $I_0 \propto T^{-\gamma}$ 
with an edge exponent $\gamma = 0.43 \pm 0.05$. Such a strong temperature
dependence is characteristic for a FES and allows 
us to exclude that pure density of states effects in the 3D emitter
are responsible for the current peaks in high magnetic fields.
\begin{figure}
\vspace*{1em}
\centerline{\epsfxsize=7cm 
\epsfbox{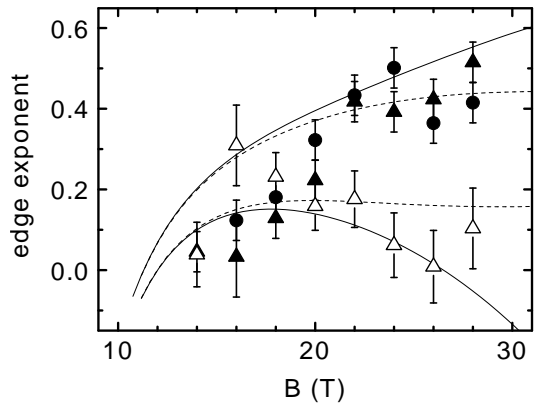}} 
\vspace*{2em} 
\caption{Experimental edge exponents $\gamma$ 
extracted from the temperature dependence of the peak height (circles) 
and from fitting the slope (triangles) compared to our theoretical predictions.
The majority spin in the emitter is shown with filled symbols, 
open symbols correspond to the minority spin. The solid lines represent the 
theoretical prediction without level broadening, for the dashed curves
a level broadening $\Gamma = 1.3$~meV was included in the theory.}  
\label{gammas}
\vspace*{1em}
\end{figure}
As shown in Fig.~\ref{Tabh} an edge exponent $\gamma = 0.43$ also
fits within experimental accuracy the observed decrease of the current
for $V>V_0$.  

It is not possible to extract the edge exponent for the minority spin 
directly from temperature dependent experiments.
At high magnetic fields  
the observed increase of the current with increasing temperature is 
mainly caused by an additional thermal population of the minority
spin in the emitter. The general form of the curve is merely affected by
temperature. Therefore, the edge exponent can only be gained 
from fitting the shape of the current peaks.

A compilation of the edge exponents $\gamma$ for various magnetic fields
and both spin orientations is shown in Fig.~\ref{gammas}.

For the data related to the majority spin two independent methods were used 
to extract $\gamma$. For the minority spin only fitting of the
shape of the $I$-$V$-curves was used.

For a theoretical description of these effects we consider a 3D
electron gas in the half space $z<0$.  In a sufficiently strong
magnetic field $B||\hat{z}$ all electrons are in the lowest Landau
level.  This defines a set of one-dimensional channels with momentum
$\hbar k$ perpendicular to the boundary.  This situation is different
from the cases considered for scattering off point defects as in
Refs.~\cite{theoXray,Matveev:1992} or for a 2D electron gas where 
the current is carried by edge states \cite{BaMa95}.  
The single particle wave
functions in channel $m\ge0$ are $\psi_m(\rho,\phi) \sin kz$ with
$\psi_m(\rho,\phi) \propto \rho^m \exp(-im\phi-\rho^2/4\ell_0^2)$.  
In the experiments the magnetic length $\ell_0=\sqrt{\hbar/eB}$ 
($\ell_0 = 5.6$~nm at 20~T) is comparable to the lateral 
size of the QD $2 r_0 \approx 7$~nm.
Hence the effect of the electrostatic potential of a charged dot on
the electrons in a given channel of the emitter decreases rapidly with
$m$, and the observed FES are mainly due to tunneling of electrons from the
$m=0$ channel into the dot.  Following \cite{theoXray,Matveev:1992} tunneling
processes of spin $\sigma$ electrons from the $m=0$ state in the
emitter give rise to a FES with edge exponent
\begin{equation}
  \gamma_\sigma = -\frac{2}{\pi}\delta_0(k_{F\sigma})
        -  \frac{1}{\pi^2}\sum_{m}\sum_{\tau=\uparrow,\downarrow}
                 \left(\delta_m(k_{F\tau})\right)^2
\label{expo}
\end{equation}
where $\delta_m(k)$ is the Fermi phase shift experienced by the
electrons in the $m$-th channel due to the potential of the quantum
dot \cite{contact}.
From (\ref{expo}) the observed field dependence of the edge exponents
is a consequence of the variation of the Fermi momenta for
spin-$\sigma$ electrons with magnetic field \emph{and} the field dependence of
the effective potential in the one-dimensional channels.  The former
can be computed from the one-dimensional density of states (DOS) of
the lowest Landau band
\begin{equation}
  D(E,B) = \frac{e\sqrt{m^{*}}}{(2\pi\hbar)^2}\,B \left(
        d(\epsilon_\uparrow) +d(\epsilon_\downarrow) \right)\ .
\end{equation}
Here $\epsilon_\sigma = E-(\hbar\omega_c\pm g^*\mu_BB)/2$ is the energy
of electrons with spin-$\sigma$ measured from the bottom of the Landau
band. $g^* \approx -0.33$~\cite{Pfeffer:1985} is the effective
Land\'e factor of the electrons in the emitter.
The DOS for the spin-subbands is $d(\epsilon) = \sqrt{2}
{\mathrm{Re}} (\epsilon+ i\Gamma)^{-{1/2}}$.  Without broadening, $\Gamma=0$, 
one has $k_{F\sigma} = \pi^2n \ell_0^2 (1\pm b^3)$ where $n$ is the 3D 
density of electrons and $b$ is the magnetic field measured in units of the 
field necessary for complete spin polarization of the 3D emitter. Using a 
Fermi energy $E_0=13.6$~meV and neglecting level broadening we find that 
only the lowest Landau level (\emph{both spin states!}) is occupied for 
$B_1 >5.2$~T. Including level broadening changes $B_1$ to a slightly higher 
value. With the known field dependence of the Fermi energy in the quantum 
limit we can calculate the field for total spin polarisation 
\begin{equation}
 B_{pol} = \left( \frac{16}{9 \xi}\right) ^{1/3} \frac{m^* E_0}{\hbar e} 
\simeq 43~\mbox{T}
\end{equation}
with $g^*= -0.33$~\cite{Pfeffer:1985} and $m^*=0.067\,m_0$.
$\xi = \frac{1}{2} |g^*| m^*/m_0$ is the ratio between spin splitting and 
Landau level splitting.

To make contact to the experimental observations we have to specify
the interaction of the screened charge on the QD and the
conduction band electrons.  A Thomas-Fermi calculation gives
$U(\rho,z) = (2e^2 \exp(\kappa z)/\kappa) (d/(\rho^2 +d^2)^{(3/2)})$
\cite{Matveev:1992}.  Here $d=5\,\mathrm{nm}$ is the width of the insulating
layer and $\kappa^{-1}=7\,\mathrm{nm}$ is the Debye radius.  The
effective potential seen by electrons in channel $m$ is
$V_m\exp(\kappa z)/\kappa$ with $V_m = 2e^2d \int d\rho^2
|\psi_m(\rho,\phi)|^2/(\rho^2 +d^2)^{(3/2)}$. For large $\kappa$ 
we obtain for the phase shift in the $m=0$ channel
$\delta_0(k) \approx -v_0f(B)k/\kappa$ where
\begin{equation}
  f(B) = \left(\frac{d}{\ell_0}\right)^2
        \left\{ 1 - \sqrt{\pi\over2}\, {d\over\ell_0} {\rm
        e}^{d^2\over2\ell_0^2}
        \mathrm{erfc}\left({d\over\sqrt{2}\ell_0}\right)
        \right\}\
\end{equation}
and $v_0 \sim (m^*e^2/\hbar^2\kappa) (\kappa d)^{-2}$ up to a numerical factor.
Similarly we obtain the integrated effect of the channels $m>0$ in
(\ref{expo}).  In Fig.~4 the resulting exponents $\gamma_\sigma$
obtained for $\sigma=\uparrow,\downarrow$ are shown for $v_0= 6.75$ and
a broadening $\Gamma= 0$ and $\Gamma = 1.3$~meV, respectively. 
The value used for $\Gamma$ reflects its realistic experimental value. 
$v_0$ is the only fit parameter.

Already the simple model with no level broadening ($\Gamma = 0$) is in good 
agreement with the experimentally measured edge exponents for both spin 
directions, especially in high magnetic fields where possible admixtures of 
higher Landau levels play a minor rule. Including level broadening leads to a 
less dramatic spin polarisation in the emitter and as a consequence smears 
out the field dependence of $\gamma$ for the minority spin. The basic 
features, however, remain unchanged. In particular, the edge exponent for the 
minority spin retains moderate values for high magnetic fields, whereas the 
edge exponent related to the majority spin shows a strong field dependence 
with very high values in high magnetic fields.  

In conclusion we have evaluated experimental data concerning 
magnetic-field-induced FES in resonant 
tunneling experiments through InAs QDs. 
We have shown that the interaction between a localized charge
and the electrons in the Landau quantized emitter leads to
dramatic Fermi phase shifts if only the lowest Landau
level in the 3D emitter is occupied.
This results in edge exponents $\gamma > 0.5$ which were measured and
described theoretically.

We would like to thank H.~Marx for sample growing, 
P.~K\"onig for experimental support and F.~J.~Ahlers for valuable discussions.
Part of this work has been supported by the TMR Programme of the European 
Union under contract no.~ERBFMGECT950077.
We acknowledge partial support from the Deutsche
Forschungsgemeinschaft under Grants HA~1826/5-1 and Fr~737/3.


\begin{references}
\bibitem{explain-alpha}
The energy-to-voltage conversion
factor $\alpha$ is derived from the temperature dependence of
the width $\delta  V$ of a current step caused by the thermal smearing
of the Fermi-edge in the emitter.   
Only a part $\alpha$ 
of the total voltage applied drops between the emitter and the dot, 
the rest of the voltage drop occurs inside the electrodes and between
dot and collector.
\bibitem{contact} In fact, one has to consider the change of this
phase shift due to the increase in charge on the QD here.
For the contact interaction considered here this is not essential.
\end{references}
\end{document}